\documentclass[aps,prb,twocolumn,hidepacs,hidekeys,superscriptaddress,10pt]{revtex4}

\newcommand{\comment}[1]{}

\usepackage{amsmath}
\usepackage{amssymb}
\usepackage{latexsym}
\usepackage{indentfirst}
\usepackage{graphicx}
\usepackage{color}
\usepackage{bm}
\usepackage{multirow}

\begin{document}

\title{Asymmetric switching behavior in perpendicularly magnetized spin-valve nanopillars due to the polarizer dipole field}

\author{D.~B. Gopman}
\author{D. Bedau}
\email{db137@nyu.edu}
\affiliation{Department of Physics, New York University, New
             York, NY 10003, USA}
\author{S. Mangin}
\author{C.~H. Lambert}
\affiliation{Institute Jean Lamour, UMR CNRS 7198, Nancy Universit\'{e},
             Vandoeuvre, France}
\author{E.~E. Fullerton}
\affiliation{CMRR, University of California at San Diego,
             La Jolla, CA 92093, USA}
\author{J.~A. Katine}
\affiliation{San Jose Research Center, Hitachi-GST,
             San Jose, CA 95135, USA}
\author{A.~D. Kent}
\affiliation{Department of Physics, New York University, New
             York, NY 10003, USA}

\begin{abstract}
We report the free layer switching field distributions of spin-valve nanopillars with perpendicular magnetization. While the distributions are consistent with a thermal activation model, they show a strong asymmetry between the parallel to antiparallel and the reverse transition, with energy barriers more than 50~\% higher for the parallel to antiparallel transitions. The inhomogeneous dipolar field from the polarizer is demonstrated to be at the origin of this symmetry breaking. Interestingly, the symmetry is restored for devices with a lithographically defined notch pair removed from the midpoint of the pillar cross-section along the ellipse long axis. These results have important implications for the thermal stability of perpendicular magnetized MRAM bit cells.
\end{abstract}


\maketitle

Magnetization reversal in magnetic nanostructures has been studied extensively, both to gain a better understanding of the underlying magnetic interactions \cite{jS02}, as well as to optimize the energy barrier for magnetic storage applications \cite{bT2005}. Nanopillars with perpendicular magnetic anisotropy are of particular importance to MRAM applications \cite{aK2010,sI2010}. The all-perpendicular geometry yields reduced critical currents, high stability and good efficiency (e.g., small $ \text{I}_\text{c} $/U) \cite{jS00,sM06,sM09}. This geometry also gives rise to an out-of-plane dipole field from the polarizer, which can shift the center of the free layer minor hysteresis loop by a considerable fraction of the room temperature coercive field. 

Generally the interactions between layers in a spin-valve are described by a single dipole field. This consideration is sufficient to understand the shift of the center of free layer minor hysteresis loops and is relevant to determine regions of bistability for spintronic applications. However, the field from a uniformly magnetized polarizer at the height of the free layer may vary by more than 100~\% between the center and the edges of a spin-valve nanopillar. The inhomogeneity of the polarizer field may play a larger role in the reversal of the free layer than can be described by a single dipole field. This can be addressed through study of the statistics of magnetization reversal for different transitions of the free layer with respect to a fixed polarizing layer magnetization.

In this letter, we present switching field distributions for switching the free layer in all-perpendicular magnetized spin-valve nanopillars. We measure the probability of switching as a function of applied fields under a linearly ramped external field. The resultant switching field distributions are strongly dependent upon the rate of thermally activated magnetization switching, which depends sensitively on the energy barriers separating metastable and stable magnetization states. We demonstrate that switching field distributions are a powerful tool to investigate the reversal transitions of a nanomagnet and show the influence of sample geometry and orientation of the polarizer magnetization in all-perpendicular spin-valve nanopillars.

\begin{figure}[!t]
  \begin{center}
    \includegraphics[width=3.in,
    keepaspectratio=True]
    {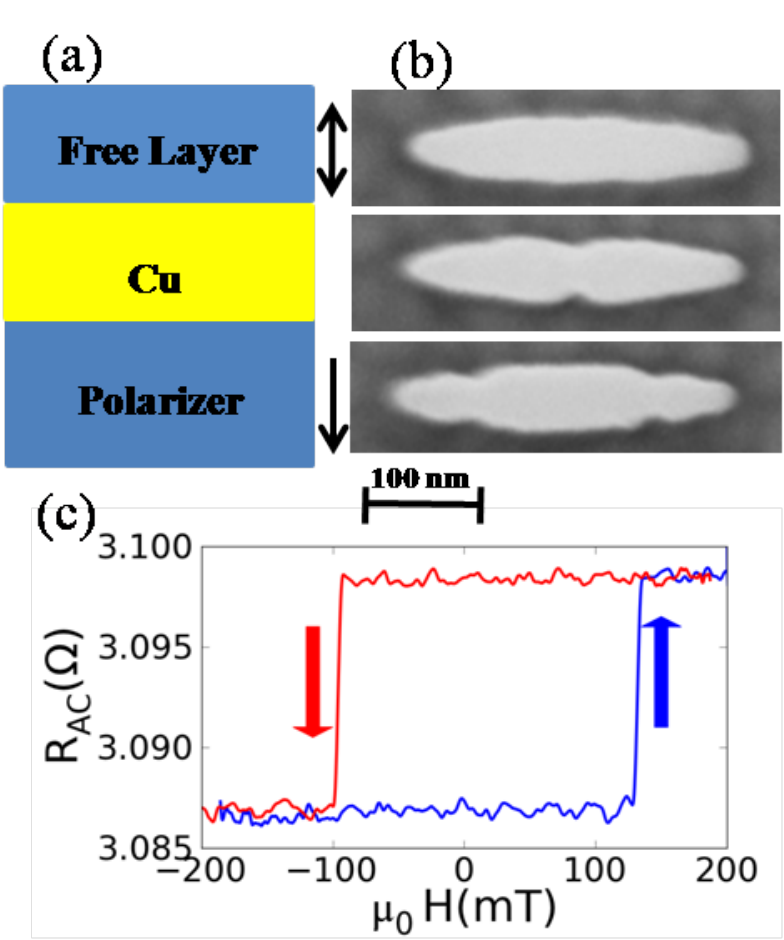}
  \end{center}
  \caption{\label{fig:SCHEMATIC} (a) Schematic of a spin-valve nanopillar; (b) SEM image of  pillar cross-section for $50~\times~300~\text{nm}^2$ ellipses showing no notch, a single notch and double notch pairs; (c) typical hysteresis loop for sample with no notch.}
\end{figure}

Our spin-valve nanopillars are magnetic multilayered films with strong uniaxial anisotropy perpendicular to the plane and have been described previously \cite{sM06,dB10a,dB10b}. The free layer is a Co/Ni multilayer and the polarizing layer is a Co/Ni Co/Pt multilayer with a sufficiently higher coercive field to be considered fixed for all of our measurements (see Fig.~\ref{fig:SCHEMATIC}(a)). The layer stack is comprised of Ta(5)\slash Cu(30)\slash Pt(3)\slash [Co(0.25)\slash Pt(0.52)]$\times$4\slash Co(0.25)\slash [Ni(0.6)\slash Co(0.1)]$\times$2\slash Cu(4)\slash [Co(0.1)\slash Ni(0.6)]$\times$2\slash Co(0.2)\slash Pt(3)\slash Cu(20)\slash Ta(5) (layer thicknesses in nanometers). These films have been patterned into  $50~\times~300~\text{nm}^2$ ellipses by a process that combines e-beam and optical lithography. We will present results on pillars with zero, one or two lithographically defined notch pairs on each side of the long axis of the ellipse, as shown in Fig. 1(b).

Quasi-static measurements of the sample magnetoresistance were taken at room temperature using a lock-in detection scheme, with a 10~kHz excitation current of $I_{ac}=100\; \mu $A~rms (the room temperature, zero-field switching current, $ I_c \approx~5~\text{mA} \gg I_{ac} $). Minor loops of the free layer were recorded using a linear ramped magnetic field with a constant rate $v$ = 100 mT/s, as shown in Fig.~\ref{fig:SCHEMATIC}(c). Our samples have a room temperature coercivity $\mu_0H_c\approx$~100~mT and an average dipole field $\mu_0H_D\approx$~20~mT, which is defined as the shift of the center hysteresis loop at room temperature. Switching field distributions were acquired by ramping the applied field many times and recording the total field at which the free layer reverses, defined by the field at which there is a step change in sample resistance. For each sample studied here, we have recorded over 10,000 switching events for switching parallel (P) and antiparallel (AP) to the fixed reference layer.

\begin{figure}[t]
  \begin{center}
    \includegraphics[width=3.0in,
    keepaspectratio=True]
    {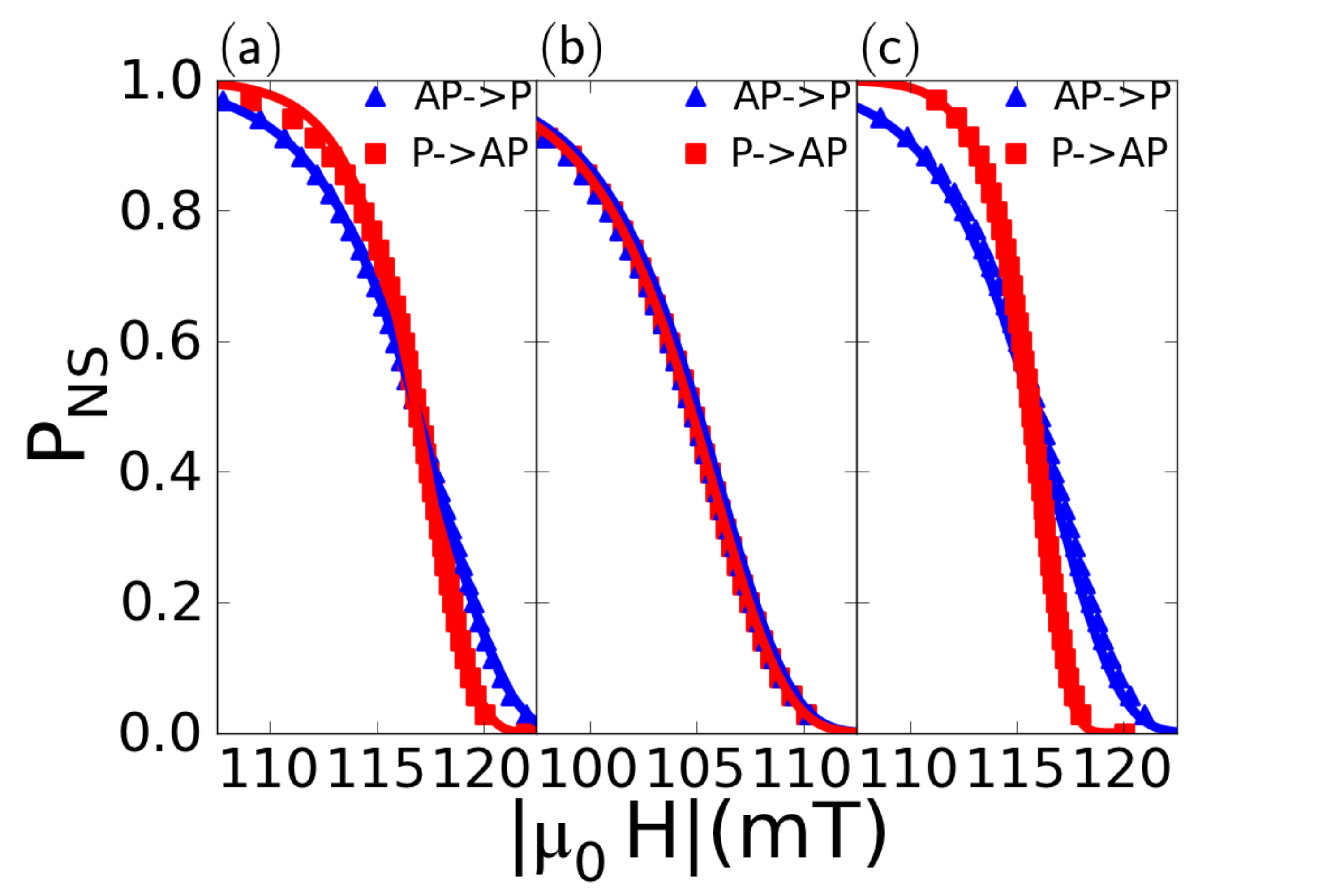}
  \end{center}
  \caption{\label{fig:SWITCHING-DISTS} The survival function $P_{NS}$ for the AP and P states for a $50~\times~300~\text{nm}^2$ junction with (a) no notch; (b) one notch pair in the center and (c) double notch pairs on each side. Data is measured against total field, defined as applied field plus a dipole field from the polarizer. AP$\rightarrow$P data (for sweeping in the negative direction) has been reflected about zero field to compare with the P$\rightarrow$AP data. Symbols are measured data and solid lines are calculated from the Kurkij\"{a}rvi model.}
\end{figure}

The survival function $P_{NS}$ for the AP and P states are displayed in Fig.~\ref{fig:SWITCHING-DISTS}. All data sets are plotted against the absolute value of the net field defined as, $H$ = $H_\mathrm{app}-H_D$. All data sets intersect around 50~\% probability, which is by definition of the dipole field $H_D$. We note that for these measurements we should expect identical switching fields for switching parallel or antiparallel to the polarizer, except for a constant shift due to the dipole field of the polarizer. However, the AP$\rightarrow$P switching field distributions are consistently wider than P$\rightarrow$AP distributions in all devices except for the single notch pair sample. This effect only depends on the relative orientation between the layers, as we have observed that reversal of the polarizer preserves the asymmetry between AP and P states (see the table below). We will examine the distributions more closely by fitting the cumulative distributions to a thermal activation model.

At nonzero temperatures, there is a finite probability that magnetization reversal takes place by thermal activation over a field-dependent energy barrier. Assuming thermal activation over a single energy barrier, at fixed temperatures and fields one can define a rate of escape as $\Gamma(H) = \Gamma_0 e^{-\beta E ( H ) }$, where $\Gamma_0$ is the attempt frequency, $\beta = 1/k_BT$ and $k_B$ is Boltzmann's constant. The form of the energy barrier, $E(H)$ is E(H) = $E_0(1-H/H_{c0})^\beta$, where $\beta$=1.5 \cite{wC95,beta}, $E_0$ is the energy barrier at zero field, and $H_{c0}$ is the zero-temperature coercive field \cite{rV89}. The cumulative probability to remain in a metastable magnetization state under finite field, $\mu_0H$, is $\exp{[-\frac{1}{v}\int_0^{H}\,\Gamma(H')dH']}$, where $v$ is the ramp rate of the magnetic field. This expression for the probability of not switching the magnetization at finite field has been attributed to Kurkij\"{a}rvi for superconductors and will be termed the Kurkij\"{a}rvi model \cite{jK72,aG95}. Figure~\ref{fig:SWITCHING-DISTS} portrays the Kurkij\"{a}rvi fits to switching field distributions as a function of total applied field. 

Consistent with the differences in the measured switching field distributions, the energy barriers that we have extracted from fitting the distributions for $\Gamma_0$ = 1~GHz and $v$ = 100 mT/s are dissimilar for all transitions except for the single notch pair junction. The best-fit parameters $E_0$ and $H_{c0}$ are listed for all samples in Table \ref{Tab:FitParams}. Most striking is the dissimilarity between energy barriers for the double notch pair pillar of 67~$k_BT$ for AP$\rightarrow$P and 145~$k_BT$ for P$\rightarrow$AP transitions ($T=300$~K).

Magnetization switching by coherent rotation should only result in a shift in the coercive field by an amount equal to the averaged polarizer field across the free layer. However, it has been indirectly observed \cite{jC09,dB10a,sK09} and more recently confirmed with simulations and imaging that magnetization reversal in thin-film nanomagnets (cross-sectional areas $\gtrsim 50^2$~nm$^2$) proceeds by sub-volume nucleation followed by the propagation of domain walls \cite{dB11,jS11}. Our results indicate that the magnitude and direction of the polarizer dipole field influences the region where magnetization reversal is initiated.

\begin{figure}[h]
  \begin{center}
    \includegraphics[width=3.0in,
    keepaspectratio=True]
   {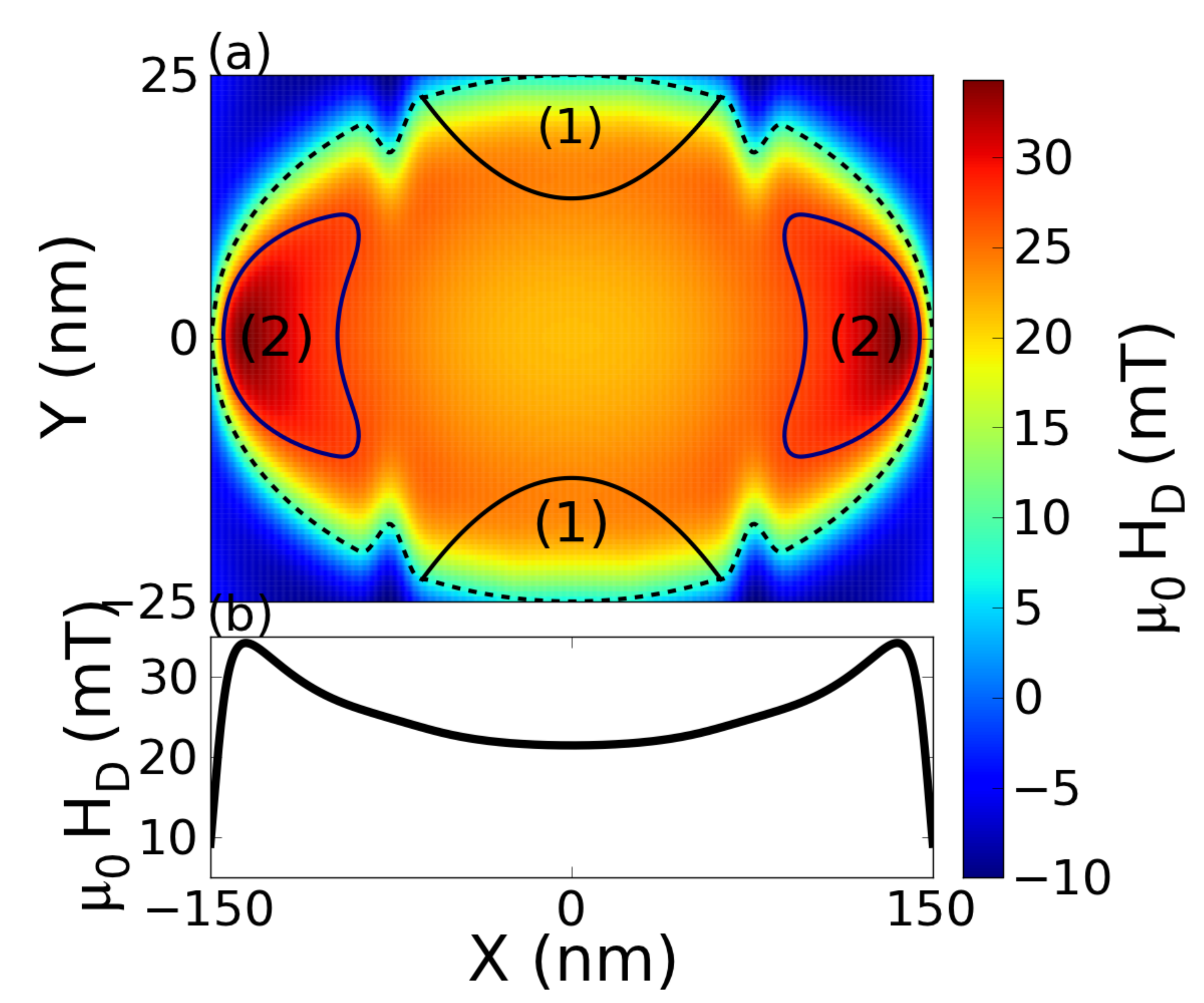}
  \end{center}
  \caption{\label{fig:FRINGE-FIELD} (a) Out-of-plane polarizer field at 5 nm above the polarizing layer for a device with two pairs of notches. Numbers (1) and (2) indicate preferred nucleation regions where the gradient of the polarizer field is low and high, respectively. (b) $y=0$ scan of polarizer field}
\end{figure}

We propose a simple interpretative model for the asymmetry in reversal starting from AP or P states. Starting from the antiparallel state, the inhomogeneous field assists in nucleation at the far edges going from AP$\rightarrow$P. Then going from P$\rightarrow$AP, the large opposing field at the edges increases the barrier for nucleation at the edges, and therefore nucleation takes place along the edges of the device center, where the polarizer fields are smallest. The relatively smaller energy barrier obtained for AP$\rightarrow$P is a consequence of distinct reversal processes for AP$\rightarrow$P and P$\rightarrow$AP transitions. Figure~\ref{fig:FRINGE-FIELD} illustrates the nucleation regions and gradient in the polarizer field along the free layer.

\begin{table}[h]\footnotesize
	\noindent{}\begin{tabular}{cc|c||c||c|l}	
	\cline{3-5}
	& & No Notch & Single Notch & Double Notch  \\ \cline{1-5}
	\multicolumn{1}{|c|}{\multirow{2}{*}{$E_0(k_BT)$}} &
	\multicolumn{1}{|c|}{AP$\rightarrow$P} & 65 & 56 & 67 (70)& \\ \cline{2-5}
	\multicolumn{1}{|c|}{}                        &
	\multicolumn{1}{|c|}{P$\rightarrow$AP} & 96 & 56 & 145 (142) & \\ \cline{1-5}
	\multicolumn{1}{|c|}{\multirow{2}{*}{$| \mu_0 H_{c0}|$(mT)}} &
	\multicolumn{1}{|c|}{AP$\rightarrow$P} & 200 & 194 & 195 (194) &    \\ \cline{2-5}
	\multicolumn{1}{|c|}{}                        &
	\multicolumn{1}{|c|}{P$\rightarrow$AP} & 170 & 194 & 151 (154) &    \\ \cline{1-5}
	\end{tabular}
\caption{Barrier heights and coercive fields for AP$\rightarrow$P, P$\rightarrow$AP transitions for different lateral geometries. Data in parenthesis in the double notch column reflects experiments on a double notch pair sample after reversing the direction of polarizer magnetization. Reversal of polarizer in double notch pair structure maintains the asymmetry in the energetics and coercivity. }
\label{Tab:FitParams}
\end{table}

The effect of the lateral geometry can also be considered within this model. The notches may serve as preferential nucleation sites in certain cases. For the sample with a single pair of notches, it may be assumed that the free layer always nucleates a reversed domain at the notches, where the gradient of the field is minimal and sharp divergences in the magnetization of the free layer could introduce magnetic instabilities. The single notch pair data provides the only dataset that can be fit with a single coercive field, $H_{c0}$, which is indicative of identical nucleation regions for AP$\rightarrow$P and P$\rightarrow$AP switching, in which case the dipole field contributions to the total field are equal and opposite. Therefore, by adding and subtracting the dipole field to the AP$\rightarrow$P and P$\rightarrow$AP switching field data, respectively, the barrier heights are identical. On the other hand, the sample with two pairs of notches placed along the perimeter confirms our model that the polarizer field influences the region where reversal initiates. Evidence of the asymmetry in AP and P reversal processes is most evident in the distribution widths in Fig.~\ref{fig:SWITCHING-DISTS}(c) and barrier heights in Table \ref{Tab:FitParams}, where results for reversing the polarizer magnetization also confirms the significance of the direction of polarizer magnetization in the observed asymmetry.

The effect of dipolar fields on reversal should be relevant in any device where the polarizer field varies considerably across the free layer, giving rise to distinctive switching statistics and reversal pathways. Inhomogeneous fields could be used to stabilize or destabilize micromagnetic states in spin-valve nanostructures. Polarizer fields could also be designed to greatly increase energy barriers. Also, the position of lithographically defined notches can remove the asymmetry in magnetization reversal processes by energetically disfavoring an alternative reversal pathway. Finally, the use of a synthetic antiferromagnetic polarizing layer may help make the reversal more symmetric \cite{tI2010} or, alternatively, the polarizer layer could be unstructured or patterned on a larger scale than the free layer to render the polarizer field acting on the free layer both smaller and more uniform.

\section*{Acknowledgments}
This research was supported at NYU by NSF Grant DMR-1006575 and at UCSD by DMR-1008654 as well as the Partner University Fund (PUF) of the Embassy of France. 
\bibliographystyle{apsrev}

\begin{thebibliography}{20}
\bibitem[{bet()}]{beta}
\bibinfo{note}{The arguments presented in this paper are valid for $1 \leq
  \beta \leq 2$, and only have a minor effect on the fitting parameters.
  $\beta= 1.5$ is taken because this applies to cases in which a magnetic field
  is applied with a slight offset to the free layer easy axis (see ref.
  \cite{wC95}).}
\bibitem[{sin()}]{singleHc0}
\bibinfo{note}{The switching field distributions (except for the single notch
  pair sample) do not fit the data if one forces a single zero-temperature
  coercive field fitting parameter for both AP$\rightarrow$P and
  P$\rightarrow$AP distributions.}
\end{thebibliography}

\end{document}